\newcommand{\edition}{ed01}
\newcommand{\version}{arXiv:\ v5,\ \ \today\ (\edition)}

\documentclass[%
onecolumn,%
oneside,%
floats,%
aps,%
prd,%
nobibnotes,%
nofootinbib,%
amsmath,%
amssymb,%
amsfonts,%
superscriptaddress,%
]{revtex4}

\usepackage[utf8]{inputenc}
\usepackage{graphicx,array,dcolumn}
\usepackage{cases}
\usepackage[paperwidth=210mm,paperheight=297mm,centering,hmargin=1.8cm,vmargin=2.5cm]{geometry}
\usepackage{enumerate}

\usepackage{hyperref}

\usepackage[normalem]{ulem}
\usepackage{soul}
\usepackage{bm}
\usepackage{bold-extra}

 \def\la{\mathrel{\mathpalette\fun <}}

\def\fun#1#2{\lower3.6pt\vbox{\baselineskip0pt\lineskip.9pt
\ialign{$\mathsurround=0pt#1\hfil ##\hfil$\crcr#2\crcr\sim\crcr}}}

\newcommand{{\SD}}{\rm SD}

\newcommand{{\Mc}}{\mathcal{M}}

\def\-g{\sqrt{-g}}

\newcommand{\be}{\begin{equation}}
\newcommand{\ee}{\end{equation}}

\newcommand{\ben}{\begin{equation*}}
\newcommand{\een}{\end{equation*}}

\newcommand{\bea}{\begin{eqnarray}}
\newcommand{\eea}{\end{eqnarray}}

\renewcommand\rho{\varrho}

\renewcommand\Re{\operatorname{Re}}

\begin{document}



\title{\sc \Large{the effect of collisions with the wall on neutron-antineutron  transitions }}

\thanks{\version} 


\author{\firstname{B.O.}~\surname{Kerbikov}\medskip}

\email{borisk@itep.ru}

\affiliation{Alikhanov Institute for Theoretical and Experimental Physics,
Moscow 117218, Russia \smallskip}

\affiliation{Lebedev Physical Institute, Moscow 119991, Russia \smallskip}

\affiliation{Moscow Institute of Physics and Technology, Dolgoprudny 141700,
Moscow Region, Russia \bigskip}

\date{\today}

\begin{abstract}
\noindent We  examine what role do collisions with the wall play in neutron-antineutron transitions. The  collision time for $n$ and $\bar n$ and the phase difference between reflected $n$ and $\bar n$ waves are calculated. It is shown that transitions inside the wall are negligible. 
 
\end{abstract}

\maketitle

\large

\section{Introduction \label{intro} }

The observation of neutron transformation to antineutron would be a  discovery
of fundamental importance which would reveal the existence of  physics beyond the SM. The possibility of  $n-\bar n$ conversion  was suggested almost half a century ago \cite{1,2,3}.  This process was extensively discussed in a recent review 
paper \cite{4}, where the  interested reader may find  the description of  different theoretical and  experimental aspects of the problem as well as a  comprehensive list of  references. 
 There are  basically three 
  experimental settings aimed at the observation of  the $n - \bar n$ conversion. The first one is to use the neutron beam from a reactor or   a
spallation source.  The beam propagates a long distance (possibly within a ballistic guide) to a target in which  the  antineutron  
 component is detected. The second
one is to  establish a limit on nuclear stability because $\bar n$ produced
inside a  nucleus will blow it up. The third option is to use ultracold
neutrons (UCN) confined in a trap.

The current  limit on the  oscillation time
$\tau_{n-\bar n} > 0.86\cdot 10^8$ s was settled  long ago by the first method at the Institute Laue-Langevin (ILL) in Grenoble \cite{5}.  The  data on  nuclear stability lead under surtain model assumptions to a similar lower bound on $\tau_{n\bar n}$ \cite{4}. 
 
In all three cases the data analysis involves a variety of theoretical problems. It is beyond the scope of the present work to dwell on this subject as a whole.  We would like to highlight the two problems -- the first one  has been solved only recently, the other is  is the subject of present work. 

The first problem arises due to the fact  that both in the beam and the trap experiments one has to take into account collisions between the neutron and the residual gas molecules.  A commonplace is to treat this effect in terms of  the Fermi potential  (see \cite{4} and  references there in ). This approach is highly objectionable since in experiments with   thermal  neutrons and 
  even with UCN the  neutron   wavelength is much smaller than the average intermolecular distance of  the  residual gas \cite{6}. The  neutron ``can see'' the individual gas molecules which rules out the use of Fermi potential.  
  This is  true for ILL experiment, experiments with trapped UCN and for the planned ESS experiment \cite{7}. The density matrix formalism is an adequate tool to describe a quantum system ($n\bar n$ in our case) in a contact with the environment. 
  
  Interaction with the environment leads to  the quantum  decoherence, i.e., to the  downfall of the  off-diagonal elements of the density matrix.  One should keep in mind that the lower bound $\tau_{n\bar n} > 10^8$s means an extremely small value of the mixing parameter  $\varepsilon=\tau^{-1}_{{n}\bar{n}}\lesssim10^{-23}$ eV. As a result the  `` quantum friction'' caused by the  interaction with the  environment leads to a  drastic   modification of the $n\bar n$  conversion law. The possible $n\bar n$ transition  process proceeds in the overdamping instead of the  oscilation regime -- see \cite{6} for a detailed description.
  
  The  second problem  which  constitutes the subject of the present work is the role  of the collisions with the guide (the trap) walls in the process of $n\bar n$  conversion. The simplest assumption which is not correct is that each collision results in  a complete randomization so that the quadratic free-space time dependence is interrupted at each collision. The idea that it  might be  possible to minimize the  randomization   through a suitable choice of the  wall material, trap size, or  reflection angle was put  forward long ago \cite{8,9}. Since then the problem was at the focus of several publications \cite{10,11,12} but despite some progress it  is still pending. In the  present work we try to analyse  collisions with the wall from  different angles. We find this work well timed in view  of the new experiments in preparation aimed at the discovery of the $n\bar n$ conversion \cite{7,13}.
  
  The paper is structured as follows. In Section II we  calculate the   reflection coefficients for  the neutrons and the antineutrons using  an  energy independent   optical potential of the wall and assuming the reflection to be  instantaneous. In Section III the wave packet formalism is  developed which allows to determine the collision time which may be quite different for $n$ and $\bar n$. Section IV is devoted to the  time-dependent approach to  the processes inside the penetration depth of the wall. We solve a coupled-channel  problem with two different initial conditions. In one case the system has some admixture of the antineutrons when it hits the wall. In the second case a pure neutron beam hits the wall and we calculate possible antineutron component created inside the wall. As expected this turns out to be negligible. In Section V the results are summarised.
  
  \section{Reflection from the Wall: Optical Model}
  
  Interaction of $n$ and $\bar n$ with the wall will  be in this Section considered in terms of the energy independent optical potential.  Our main interest concerns strongly absorptive interaction of the $\bar n$ component with the wall. Very weak absorption of the $n$ component will be ignored. Neutrons with energies $E< U_n ( U_n \sim 10^{-7}$ eV is the optical potential for neutrons) undergo a complete reflection (Ya. B.Zeldovich, 1959 \cite{14}).  The  above inequality is sometimes taken as a definition of UCN. The optical potential for antineutrons $U_{\bar n} = V_{\bar n} - i W_{\bar n}$ includes  the   imaginary part responsible for the annihilation. We shall assume that $E < V_{\bar n}$ as well.  As an example below we present the values of the optical potential for $Cu$ \cite{15,16}.
  \be U_n = 1.67 \cdot 10^{-7}~{\rm eV}, ~~ U_{\bar n} = (1.16 - i 0.11)\cdot  10^{-7} {\rm eV}.\label{4}\ee
  
  The value (\ref{4}) for $U_{\bar n}$ corresponds to the scattering lengths \cite{16}, in our earlier work \cite{11} there was an error in the value of $U_{\bar n}$. In \cite{13} we find a value $U_{\bar n} (Cu) = (2.2-i 0.1) \cdot 10^{-7}$ eV.
  Comparing this value with $U_n$ given by (\ref{4}) we find that there is an energy ``window'' $ (1.67<E<2.2 ) \cdot 10^{-7}$ eV ``transparent'' for $n$ but at least partly reflecting for $\bar n$, i.e.,  a kind of a filter for $\bar n$.
  
At this point we would like to stress that the antineutron-nucleus interaction is a complicated process depending on a lots of factors starting from the $\bar{n}$ velocity. Our understanding of the annihilation process is rather limited. The elementary $\bar{n}N$ $(N=p,n)$ amplitude is strongly isospin and spin dependent \cite{17,18,19}. Reliable calculations of the mean free time of propagation of $\bar{n}$ through the nuclear matter are absent. In such a situation the above difference of the parameters of the optical potential is not surprising. In different models the imaginary part of the optical potential differs by almost an order of magnitude \cite{20}. An attempt has been undertaken recently \cite{21} to develop a model for  $\bar{n}{}^{12}_{6}C$ interaction. Carbon is a material which served as a target in the ILL experiment \cite{5} and will be used in the ESS project \cite{7}.  
  
  Reflection coefficients are calculated along the usual  formulae of quantum mechanics. Consider neutrons moving along the $x$ -direction from $x=-\infty$ to the well at $x=0$ for $E<U_n$
  
 \be  \psi_n(x)= \left\{\begin{array}{ll}
 e^{ikx} - A_n(k) e^{-ikx},& x<0,\\B_n(k) e^{-\kappa_n x},& x>0, \end{array}\right. \label{5}\ee where $k=\sqrt{2mE}, ~\kappa_n= \sqrt{2m (U_n -E)}$. One has 
 \be A_n (k) = e^{i\varphi_n (k)}, ~~ \varphi_n (k) = 2 ~ {\rm \arctan}~\frac{k}{\kappa_n},\label{6}\ee
 or,\be A_n (k) =\rho_n (k) e^{i\varphi'_n (k)}, ~~ \rho_n (k) =1,~~ \varphi'_n(k) ={\rm \arctan}~ \frac{2 k\kappa_n}{\kappa_n^2 -k^2}. \label{7}\ee
 
 For the antineutron component with $E< V_{\bar n}$ similar equations read 
  
 \be  \psi_{\bar n}=(k) \left\{\begin{array}{ll}
 e^{ikx} - A_{\bar n}(k) e^{-ikx},& x<0,\\B_{\bar n}(k) e^{-\kappa_{\bar n} x},& x>0, \end{array}\right. \label{8}\ee
 where $ \kappa_{\bar n} = \kappa'_{\bar n} - i \kappa''_{\bar n} = [2m (V_{\bar n}- i W_{\bar n}-E)]^{1/2}$.  One gets 
 \be A_{\bar n} (k) =\frac{\kappa_{\bar n} + ik}{\kappa_{\bar n} -ik} = \rho_{\bar n}(k) e^{i\varphi_{\bar n} (k)}, \label{9}\ee
 \be \rho^2_{\bar n} (k) = \frac{(\kappa''_{\bar n} - k)^2 + \kappa'^2_{\bar n}}{(\kappa''_{\bar n} + k)^2 + \kappa'^2_{\bar n}}, ~~ \varphi_{\bar n} (k) = \arctan \frac{2k\kappa'_{\bar n}}{\kappa'^2_{\bar n} + \kappa''^{2}_{\bar n} - k^2}.\label{10}\ee
 Note that for $\kappa''_{\bar n} =0$ equations (\ref{7}) and (\ref{9}) -(\ref{10}) coincide. One recognizes in $\rho^2_{\bar n} (k)$ the familiar Fresnel  reflection coefficient. The quantity $\rho^2_{\bar{n}}$ characterizes the amount of the reflected antineutrons. In view of the uncertainty of the value of $k''_{\bar{n}}$ (or $W_{\bar{n}}$) one has at best only educated guesses on $\rho^2_{\bar{n}}$ for different materials. 
 
 An important physical quantity is  the phase difference $\theta(k) = \varphi_{\bar n} (k) - \varphi'_n (k)$ An  appealing  approximation is to consider  $\kappa''_{\bar n} << \kappa'_{\bar n}$ and $k<< \kappa_n, ~~ \kappa'_{\bar n}$. Then 
 \be \theta(k) \simeq \frac{ 2k}{\kappa_n \kappa'_{\bar n}} (\kappa'_{\bar n} - \kappa_n).\label{11}\ee
 
 One may argue that at $k\to 0$ the phase  difference $\theta(k)$ may be discarded. Certain caution over this conclusion is needed. The first trivial remark is that particles hitting the wall form a wave packet. For UCN the typical energy resolution is  $\Delta E/E \sim 10^{-3}$ \cite{10,11}
 which  means  that the wave packet contains a band of momenta $\Delta k \sim 10^{-2}$ eV
 around the central value. This is turn results in the variations of  $\theta (k$). 
What is more important, is that equations (\ref{5})-(\ref{6}) and (\ref{8})-(\ref{9})
do not contain the $n-\bar n$ coupling. In the correct equations the waves hitting the wall and reflected from 
it must be mixtures of the $n$ and $\bar n$ components. For the conversion inside nuclei coupled equations have been 
formulated and  solved using either the optical potential \cite{20,22,23,24,25} or the Bloch equations \cite{6}.
To our  knowledge Ref. \cite{11} is the only paper treating the interaction with the surface of the  wall as a 
coupled channel problem.
However,the final calculations in \cite{11} were  performed under the assumptions  that the  interaction with the surface is  instantaneous.

Even in this approximation the influence of the phase difference $\theta$ is very strong. Calculations \cite{11} performed at
 $E=0.8 \cdot 10^{-7}$ eV $(v =3.9\cdot 10^2$cm/s) for Cu show that the  account of the phase difference leads after 10 collisions to a suppression 
 factor $\simeq 5$ for the $n\bar n$  conversion probability \cite{11}.

In reality the collision time for $n$ and $\bar n$ may  be very different -- see below.
 It will be shown that the neutron collision time $\tau_n$ may  diverge 
(at least formally) in the limits $E\to 0$ and $E\to U_n$ from below.

The collision time $\tau_{\bar n}$ for the  antineutron is bounded from above by the annihilation
time $1/\Gamma$ inside the wall. To determine the collision time one has to describe the interaction with the wall in terms of the  wave packets.

\section{The wave Packet Approach. Collision time}

The  wave packet formalism is  at  the core of the time dependent description of the  quantum collision process. The wave packet for a single
 particle $(n, $ or $ \bar n$)  may be  written  in the form 
\be \Psi (x,t)   = \frac{N}{2\Delta E} \int^{E+\Delta E}_{E-\Delta E} d E'  \Psi_{E'} (x,t), \label{13}\ee
where $N$ is a  normalization constant and  $ \frac{d^2 k}{dE^2} \Delta E \ll \frac{dk}{dE}$, or $\Delta E \ll E$.
Recalling equation (\ref{5}) for the  neutron  reflected wave we write

\be \Psi (x,t)   = \frac{N}{2\Delta E} \int^{E+\Delta E}_{E-\Delta E} d E' \left(e^{ik'x}-A_n(E') e^{-ik'x}\right) e^{-iE't}.\label{151}\ee

Next we expand $k'(E')$ and $A_n (E')$ in the vicinity of $E'=E$ up to linear terms. This yields $(\nu=E'-E)$

\be \Psi (x,t)  = \frac{N}{2\Delta E} \int^{\Delta E}_{ -\Delta E} d \nu
 \left( e^{ikx + i \frac{dk}{dE} \nu x} - \left( A_n + \frac{dA_n}{dE}\nu \right) e^{-ikx-i\frac{dk}
{dE}\nu x}\right)e^{-iEt-i\nu t}.\label{152}\ee

The result of the integration reads
\be \Psi(x,t) =N \left[\mu (\xi) e^{ikx-iEt} -
\frac{\kappa_n + ik}{\kappa_n -ik} \left( \mu (\eta) - \eta_0 \frac{ d \mu(\eta)}{d\eta} \right) e^{-ikx-iEt} \right],\label{16}\ee
where $\mu(x) = \frac{{\rm sin}~ x}{x}, \xi = \Delta E \left( \frac{x}{v}-t\right), ~~ \eta = \Delta E \left( \frac{x}{v} + t\right),$ 
$ v^{-1} = \frac{dk}{dE}, ~~\eta_0 = \Delta E\frac{2m}{k\kappa_n}. $
For the beam resolutions $\Delta E/E \sim 10^{-3}$ the $\eta_0$ parameter is  small and (\ref{16}) can be recasted
into the form

\be \Psi(x,t) \simeq N\left[ \mu (\xi) e^{ikx-iEt}- \frac{\kappa_n +ik}{\kappa_n-ik} \mu (\eta-\eta_0 )
e^{-ikx-iEt}\right].\label{17}\ee

Equation (\ref{17}) gives the retardation (collision) time of the wave packet

\be \tau_n =  \frac{2m}{k\kappa_n} =\left[E(U_n -E)\right]^{-1/2}.\label{18}\ee

One  notices that an alternative and more familiar  way to obtain  the same result is 

\be \tau_n = \frac{ d\varphi_n}{dE} =\frac{d}{dE} \left\{2 ~{\rm \arctan}~ k/\kappa\right\} = [E(U_n -E)]^{-1/2}.\label{19}\ee

According to (\ref{18})-(\ref{19}) the  collision time increases when either $E\to 0, $ or $E\to U_n$.

Similar  derivation using the wave packet may be performed for the case of antineutrons. We remind that  according to (\ref{11})
one should replace $\kappa_n$ by a complex $\kappa_{\bar n}= \kappa'_{\bar n}-i\kappa_{\bar n}''$. The
collision time is obtained by expanding $|\varphi(\eta-\eta_0)|^2$  around $\eta= \eta_0$ and looking for a maximum

\be |\varphi(\eta-\eta_0)|^2 \simeq 1-\frac13 \eta^2+\frac23\eta \Re (\eta_0),
\label{20}\ee

\be \tau_{\bar n} = \Re \frac{2m}{k\,\kappa_{\bar n}}.\label{21}\ee

As an  illustration we present the  ratio $\gamma=\tau_n/\tau_{\bar n}$ for two different values of  $E$ and for  a set of parameters (\ref{4}). For 
 $E\to 0$ one has  $\gamma \to 0.8$,  for $E\to V_{\bar n}$ from below $\gamma \simeq 0.33$ According to (\ref{21})  the collision time for $\bar n$
  depends on the  annihilation inside the wall. In the next  Section we  develop   a time-dependent approach to the  conversion process inside the wall.

\section{Hitting the Wall. The time-Dependent Approach}

We want to describe what happens to the  $n\bar n$ system within the penetration length inside the wall. Again we assume that for the  velocities
 below the limiting value for a given material the  neutron  undergoes a complete reflection. This does not contradict the fact that the tail of the neutron wave 
function penetrates inside  the wall. On general grounds the penetration depth $l_n \sim \lambda \simeq 10^{-5}$ cm,
 with $\lambda$ being the Broglie wave length. This is in line with the wave packet formalism developed above. With $\beta$-decay   neglected the Hamiltonian of the $
n\bar n$ system inside the wall  has the form
\be H=\left( \begin{array}{ll}E_n&\varepsilon\\ \varepsilon& E_{\bar n}-i\frac{\Gamma}{2}\end{array}\right),\label{22}\ee

where $\Gamma\simeq 2 W \simeq 0.2 \cdot 10^{-7}$ eV for Cu. The two eigenvalues of this Hamiltonian are \cite{10,26}

\be \mu_1 = E_n-\left(E_\varepsilon +i \frac{\Gamma_\varepsilon}{2}\right)+...,\label{23}\ee
\be \mu_2 =E_{\bar n} -i\frac{\Gamma}{2} + \left( E_\varepsilon+ i \frac{\Gamma_\varepsilon}{2}\right)+...,\label{24}\ee

\be E_\varepsilon + i \frac{\Gamma_\varepsilon}{2} =\frac{\varepsilon^2}{\omega-i\frac{\Gamma}{2}},\label{25}\ee
where $\omega=E_{\bar n}-E_n$. In deriving (\ref{23})-(\ref{25}) expansion
over the small parameter $\varepsilon^2 |H_{11} - H_{22}|^{-2}$ has bee used. In terms of the  eigenvalues $\mu_1$ and $\mu_2$ the general solution pf the  two-component evolution equation has the form \cite{27}
\be \Psi(t) = \left( \begin{array}{l} \Psi_n (t)\\ \Psi_{\bar n} (t)\end{array} \right) = \left ( \frac{H-\mu_2}{\mu_1-\mu_2} r^{-i\mu_1t} +\frac{H-\mu_1}{\mu_2-\mu_1} r^{-i\mu_2t} \right) \Psi(0).\label{26}\ee

Consider first a solution corresponding to the following initial conditions
\be \Psi_{\bar n} (t=0)= \varepsilon \tau, ~~ \Psi_n (t=0) = \sqrt{1-\varepsilon^2\tau^2}, \label{27}\ee
where $\tau$ is some unspecified short $(\tau\varepsilon \ll 1)$ time interval during which the system propagates in a free space without ambient magnetic field \cite{4}. At $t=0$ it hits the wall and the evolution proceeds according to (\ref{26}). The  eigensolutions corresponding to the initial conditions (\ref{27}) are easily found. The result for the antineutron component reads
\be \Psi_{\bar n} (t) = \varepsilon \tau \exp \left( -iE_{\bar n} t - \frac{\Gamma}{2} t\right), ~~ |\Psi_{\bar n} (t) |^2 = \varepsilon^2 \tau^2 e^{-\Gamma t}.\label{28}\ee
This result id physically transparent. During the collision time $t$ the preformed antineutron admixture $\varepsilon^2\tau^2$ is depleted  due to annihilation. This means that the collision time $\tau_{\bar n} \la 1/\Gamma \sim 10^{-8}s$.

Another somewhat artificial initial condition is 
\be \Psi_{\bar n} (t=0)=0, ~~ \Psi_{  n} (t=0)=1.\label{29}\ee
In this case we may inquire what is the amount of antineutrons produced inside the wall during the collision. The solution has the form
\be |\Psi_{\bar n} (t)|^2 =\frac{\varepsilon^2}{\omega^2+\frac{\Gamma^2}{4}} e^{-\Gamma_\varepsilon t} \left( 1+ e^{-\Gamma't} - 2e^{-\frac{\Gamma'}{2}t} cos~ \omega t\right), \label{30}\ee 
where $\Gamma_\varepsilon$ was defined  by (\ref{25}), $ \Gamma' = \Gamma-2\Gamma_\varepsilon$. Equation (\ref{30}) contains a huge suppression factor $\varepsilon^2 (\omega^2+\Gamma^2/4)^{-2}\sim 10^{-30}.$ Therefore $n\bar n$  conversion inside the wall may be neglected. In the limit  of zero splitting $\omega =0$
\be |\Psi_{\bar n} (t)|^2 = \frac{16\varepsilon^2}{\Gamma^2} e^{-\frac{\Gamma}{2} t} sh^2 \frac{\Gamma t}{4}.\label{31}\ee

We note in passing that inside nuclei or in  nuclear matter for relevant times $\Gamma t \gg 1 (\Gamma\sim 10^{23} s^{-1})$ equation (\ref{30}) leads to a  result which looks quaite different \cite{6,26}
 
\be |\Psi_{\bar n} (t)|^2 = \frac{4\varepsilon^2}{\Gamma^2} \exp \left(-\frac{4\varepsilon^2}{\Gamma} t\right).\label{32}\ee

\section{ Conclusions and Outlook}

The problem of how to treat the interaction with the wall in the process of neutron-antineutron transitions is important both  from the theoretical point of view and also in applications   to the next  generations of experiments. In this paper  we followed three lines of arguments. The optical model for the wall  potential allows in a simple way to derive the reflection coefficients for $n$ and $\bar n$. Due to  reflection the $\bar n$ amplitude is reduced by  $\rho_{\bar n} (k)<1$ and its phase is shifted relative to the neutron component. Even  if we leave  aside our poor knowledge of the  antinucleon optical potential, the obtained  results by no means can be  considered as a  complete solution of the problem. Using the wave packet formalism these results were  extended to take into account the finite  collision (retardation) time. Depending on the strengths of the potentials and the beam energy the time  delay for $n$ and $\bar n$ may be remarkably different. In order to  shed some light on the microscopic mechanism behind the reflection coefficients and the collision time we developed the time-dependent approach to the dynamics  of the $n\bar n$ system inside the wall. The problem is solved under two different initial conditions. If the hitting beam contains some preformed antineutron admixture it is simply depleted  due to  annihilation   during the collision thus imposing an upper bound on the collision time. If one assumes that prior to the contact with the wall the beam contained only neutrons the $n\bar n$ conversion inside the wall is negligible.

One more comment on the dynamic of decoherence in the wave packet folmalism is worth making. The obvious mechanism of decoherence is the spatial separation of the $n$ and ${\bar{n}}$ wave packets reflected from the wall. Consider both packets of the same size $\sigma$ and the same velocity $v$. Let $\Delta\tau=\tau_{\bar{n}}-\tau_{n}$ be the difference of the collision times (\ref{21}) and (\ref{18}). Coherence is lost if $v\Delta{\tau} \gtrsim \sigma$. As an example take $\sigma = 10^{-4}$ cm, $v = 10^5$ cm/s, $\Delta\tau=10^{-8}$ s, then $v \Delta\tau > \sigma$ which means the lost of coherence. Less obvious source of decoherence is the irreversible deformation of the $\bar{n}$ wave packet due to annihilation. Annihilation rate is inversely proportional to the velocity. Therefore the low velocity components of the $\bar{n}$ wave packet will be depleted more intensively leading to the deformation of the wave packet shape and the damping of coherence. The qualitative description of this mechanism is beyond the scope of this paper.

Finally, we would like to mention that Eq.(\ref{30}) (which to our knowledge was first derived in \cite{10}) has been used to describe mixing between ordinary  and hidden photons: see $(2.9)$ of \cite{28} and $(13)$ of \cite{29}. Our equation (\ref{30}) differs from the two formulae mentioned above by an extra  factor $\exp{-\Gamma_{\varepsilon} t}$ which may be important in some limiting cases -- see $(27)$ of \cite{29}.   

Despite the intensive investigation performed in this article the  body of the  techniques and formulae to describe the interaction with the wall in the process of $n\bar n$ conversion is certainly not complete. The main problem for further research is to get a better understanding of the annihilation process and to combine the above approaches into a unified picture. 

I would like to thank Vadim Lensky who kindly provided the author by  his notes related to our common publication \cite{11}. In addition, I would like to thank  Valery Nesvizhevsky whose talk \cite{13} gave  the author an impetus to  this research. This work was supported by the Russian Science Foundation grant number 16-12-10414.

\end{document}